\documentclass[aps,prev,twocolumn,preprintnumbers,floatfix,nofootinbib]{revtex4-1}
\usepackage[scientific-notation=true]{siunitx}
\usepackage{float, verbatim, amsmath, amssymb, fullpage, paralist, cancel, listings, subfig, enumitem, siunitx, graphicx, xcolor, tikz, booktabs, tcolorbox, textcomp, url, parse lines, fontawesome5, marginnote, smartdiagram, forest, pifont, pgfcalendar, pgfgantt, fix-cm, ulem, soul, multirow, xfrac, colortbl, xspace, fontenc, helvet, mathptmx,ragged2e,csquotes} 
\definecolor{smcolor}{rgb}{0.8,0.5,0.5}
\definecolor{bsmcolor}{rgb}{0.4,0.6,0.8}
\definecolor{pvcolor}{rgb}{0.8,0.8,0.8}
\definecolor{cobalt}{rgb}{0.0, 0.28, 0.67}
\definecolor{darkelectricblue}{rgb}{0.33, 0.41, 0.47}
\definecolor{darkpowderblue}{rgb}{0.0, 0.2, 0.6}
\definecolor{darktangerine}{rgb}{1.0, 0.66, 0.07}
\definecolor{pastelviolet}{rgb}{0.8, 0.6, 0.79}
\definecolor{black}   {RGB}{0.0, 0.13, 0.28}
\definecolor{dukeblue}    {rgb}{0.0, 0.0, 0.61}
\definecolor{oxfordblue}{rgb}{0.0, 0.13, 0.28}
\definecolor{navyblue}{rgb}{0.0, 0.0, 0.5}
\definecolor{linkred}{RGB}{165,0,33}

\usesmartdiagramlibrary{additions}
\usepackage{hyperref}
\usepackage{cleveref}[capital]
\usepackage{textcase}
\hypersetup{
        linktocpage=true,
        setpagesize=true,
	colorlinks= true,
	linkcolor=cobalt!50!black,
        menucolor=cobalt!50!black,
	filecolor=cobalt!50!black,      
	urlcolor=cobalt!50!black,
        citecolor=cobalt!50!black,
        citebordercolor=cobalt!50!black,
	pdftitle={Light Dark Scalar},
        pdfsubject={Latex},
        pdfauthor={Yoxara Villamizar},
        pdfkeywords={BSM Phenomenology, SM, Collider, DM}
}
\graphicspath{{Figures/}} 
\newlist{firstitemize}{itemize}{1}
\setlist[firstitemize,1]{label=\color{cobalt!70!black} \ding{111}, nosep, leftmargin=*, before=\normalcolor}
\newlist{seconditemize}{itemize}{1}
\setlist[seconditemize,1]{label=\color{cobalt!70!black}\ding{226}\normalcolor, nosep, leftmargin=*} 
\newlist{thirditemize}{itemize}{1}
\setlist[thirditemize,1]{label=\color{cobalt!70!black} \ding{118}, nosep, leftmargin=*, before=\normalcolor}

\renewcommand{\bar}{\overline}

\usepackage{tikz}
\usetikzlibrary{arrows.meta,decorations,calc,shadows.blur,decorations.pathreplacing,decorations.pathmorphing,decorations.markings,shapes, arrows, positioning,bending,chains,matrix, patterns,mindmap,trees}
\tikzset{
>=triangle 45,
scalar/.style={decorate, dashed, draw=cobalt!50!black, thick, line width=0.8pt}, 
photon/.style={decorate, line width=0.85pt, draw=black, decoration={snake, segment length=5, amplitude=3pt, post length=0.5mm, pre length=0.5mm}}, 
particle/.style={draw=black, postaction={decorate}},
fermion/.style={draw=black, line width=0.6pt, postaction={decorate}, decoration={markings,mark=at position .5 with {\arrow[draw=black,scale=1.2,>=stealth]{>}}}},
antifermion/.style={draw=black, line width=0.6pt, postaction={decorate},decoration={markings,mark=at position .5 with {\arrow[draw=black,scale=1.2,>=stealth]{<}}}},
}

\usepackage{graphicx}
\usepackage{amsmath, amsthm, amssymb}
\usepackage{amsfonts}
\usepackage{subfig}
\usepackage{xspace}
\usepackage{paralist}
\usepackage{multirow}
\usepackage{paralist}
\usepackage{graphicx}
\usepackage{bm}
\usepackage{times}
\usepackage{slashed}
\usepackage{color}
\usepackage{epsfig}\usepackage{dcolumn}
\usepackage{color}
\def\br{\begin{eqnarray}}
\def\er{\end{eqnarray}}
\def\be{\begin{equation}}
\def\ee{\end{equation}}

\begin{document}

\title{LHC and HL-LHC Bounds on Visible and  Invisible Decays in the $B-L$ Model}

\author{Farinaldo S. Queiroz$^{1,2,3}$}
\author{Jilberto Zamora-Saa$^{3,4}$}
\author{Ricardo C. Silva$^{1,2,3}$}
\author{Y.M. Oviedo-Torres $^{1,3,4}$}

\email{Corresponding authors: mauricio.nitti@gmail.com}
\email{ricardo.csr@hotmail.com}

\affiliation{$^1$International Institute of Physics, Universidade Federal do Rio Grande do Norte, Campus Universit\'ario, Lagoa Nova, Natal-RN 59078-970, Brazil \\
$^2$ Departamento de F\'isica te\'orica e Experimental, Universidade Federal do Rio Grande do Norte, 59078-970, Natal, Rio Grande do Norte, Brazil\\
$^3$Millennium Institute for Subatomic Physics at High-Energy Frontier (SAPHIR), Fernandez Concha 700, Santiago, Chile. \\
$^4$Center for Theoretical and Experimental Particle Physics - CTEPP, Facultad de Ciencias Exactas, Universidad Andres Bello, Fernandez Concha 700, Santiago, Chile\\
}

\begin{abstract}
In this work, we use publicly available data from ATLAS collaboration collected at LHC run 2 at a center-of-mass energy of $\sqrt{s}=13$~TeV with an integrated luminosity of $139 fb^{-1}$ to derive lower mass limits on the $Z^\prime$ gauge boson associated with the B-L gauge symmetry.
Using dilepton data we find that $M_{Z^\prime} > 4$~TeV ($6$~TeV) for $g_{BL}=0.1$ ($g_{BL}=0.5$) in the absence of invisible decays. Once invisible decays are turned on these limits are substantially relaxed. Assuming an invisible branching ratio of $BR_{inv}=0.9$, the LHC bound is loosened  up to $M_{Z^\prime}> 4.8$~TeV for $g_{BL}=0.5$. This analysis confirms that the LHC now imposes stricter constraints than the longstanding bounds established by LEP. We also estimate the projected HL-LHC bounds that will operate with at $\sqrt{s}=14$~TeV and a planned integrated luminosity of $\mathcal{L}=3 ab^{-1}$ that will probe $Z^\prime$ masses up to $7.5$~TeV.
\end{abstract} 
    
\maketitle

\section{Introduction}
\label{introduction}

The Standard Model (SM) of Particle Physics is a successful theoretical framework that elegantly describes the strong, electromagnetic and weak interactions between elementary particles. The Higgs discovery in 2012 was a remarkable triumph in its long and successful history \cite{ATLAS:2012yve, CMS:2012qbp}. The most memorable achievements of the SM are closely related to the history and developments of high energy accelerators. In particular, the LHC physics program stands as one of the most triumphant experimental endeavors in science. Discovering the Higgs boson was simply one of its objectives. The LHC also seeks to explore a vast array of new physics scenarios that could be manifested at the TeV scale. For this reason, an upgrade LHC in many ways including its center of mass energy and luminosity has been approved \cite{Apollinari:2015wtw,CidVidal:2018eel}.

The absence of clear new physics signals so far motivates us to constrain possible TeV scale collider signatures to corner the parameter the immense parameter space where new physics may sits. One possible outcome is that new physics could be lurking at slightly higher masses or weaker couplings than previously thought, challenging its detection. 

That said, in this work we investigate an extension of the SM due to its limitations to address the matter-antimatter asymmetry of the universe \cite{Cvetic:2021itw, Cvetic:2015naa, CarcamoHernandez:2022fvl, deJesus:2023lvn, Cogollo:2023twe}, dark matter \cite{Bertone:2004pz,Alves:2013tqa,Bertone:2016nfn,Arcadi:2017kky}, neutrino oscillations and their non-zero masses  which is based on a local gauge symmetry. Knowing that baryon (B) and lepton (L) numbers are accidental global symmetries in the SM, we promote them to gauge symmetry at the expense of adding three right-handed neutrinos to cancel the gauge anomalies.  This $U(1)_{B-L}$ model has been subject of numerous beyond the SM studies tackling these open problems  \cite{Mohapatra:1979ia,Ma:1998dx, Mohapatra:1986bd}. 

The common figure in this B-L model is the massive $Z^\prime$ boson. In the B-L model the $Z^\prime$ field couples to all SM fermions with similar coupling strength. Therefore, we take advantage the publicly available data recorded by the ATLAS collaborations in proton-proton collisions with a center-of-mass energy of $\sqrt{s}=13$~TeV during Run 2 to constrain both the gauge coupling associated with this gauge symmetry, and the mass of the $Z^\prime$ boson \cite{Klasen:2016qux,Cox:2017eme,Liu:2022kid,Fiaschi:2022wgl,CMS:2023ooo}. As dilepton events offer a relatively much cleaner environment, we will the search for high-mass dielectron and dimuon resonances in the mass range of $250$~GeV to $6$~TeV as reported by the ATLAS experiment to derive our bounds. In the original B-L model, the $Z^\prime$ boson may decay into right-handed neutrinos but the branching ratio into right-handed neutrinos is naturally small because they interact with similar strength to the $Z^\prime$ boson. Interested in covering possible dark matter realizations in the B-L model \cite{Klasen:2016qux,Patra:2016ofq,Bernal:2018aon}, we obtain limits in the presence of sizeable invisible decays as well.

Moreover, having in mind the ongoing plan to build the High-Luminosity LHC (HL-LHC) \cite{Adolphsen:2022ibf} we forecast the HL-LHC sensitivity to the model for several benchmarks. In summary, we present updated collider limits on the B-L model using LHC data  in the presence or absence of invisible $Z^\prime$ decays and estimate the HL-LHC sensitivity. 

The paper is organized as follows: in the Section \ref{blmodel} we describe the B-L model; In the Section \ref{ATLASanalysis}, we describe our strategy to interpret and extract the constraints using experimental data published by the ATLAS collaboration. In sections \ref{scenario1} and (\ref{scenario2}) we present our bounds in the absence (presence) of invisible decays. In Section\ref{HLLHCsec} we estimate the HL-LHC sensitivity before drawing our conclusions in Section \ref{conclusions}.

\section{The minimal $B-L$ model}
\label{blmodel}

\begin{figure}[ht]
    \begin{minipage}[b]{0.5\textwidth}
        \includegraphics[width=0.9\textwidth]{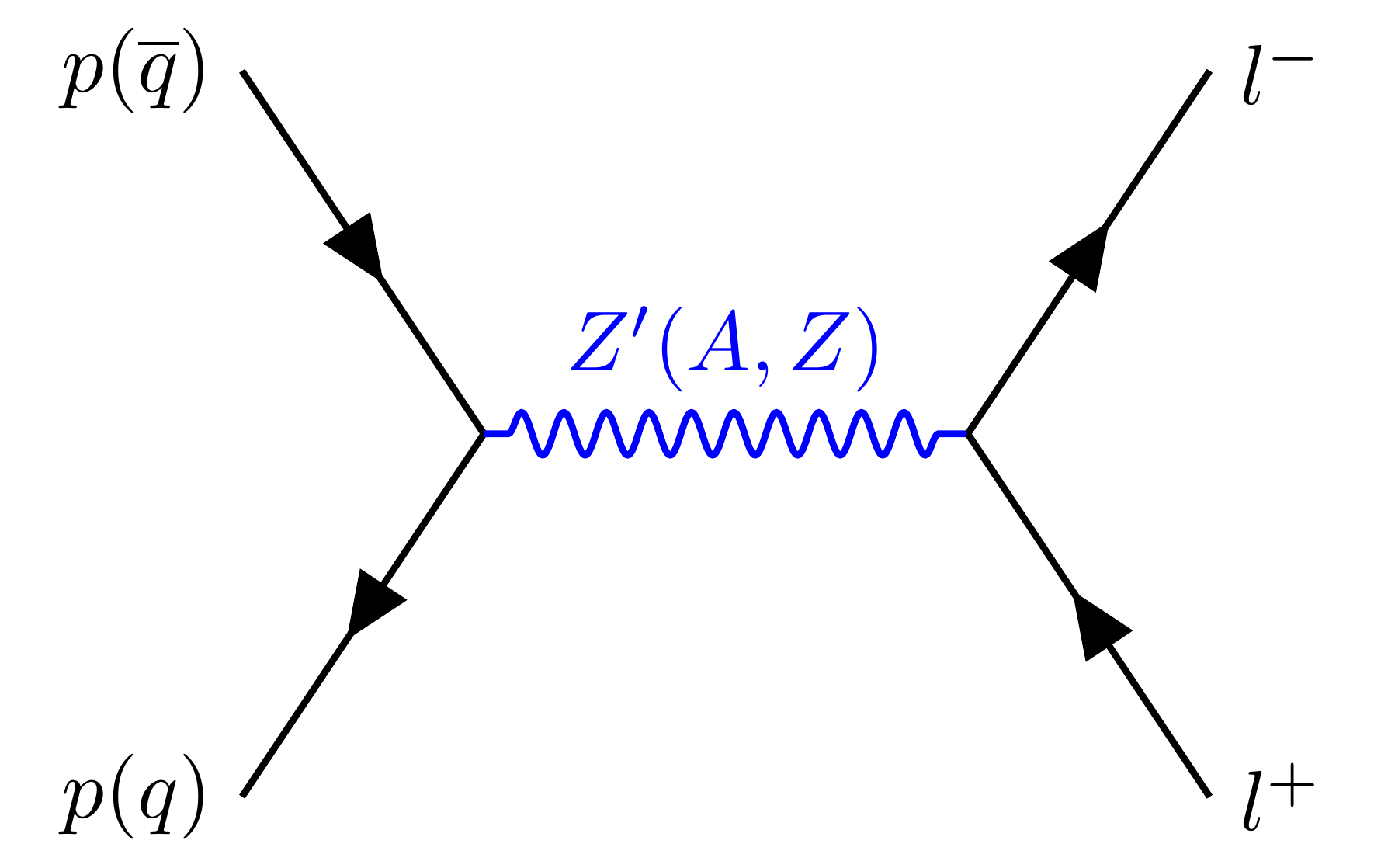}
        \caption{Feynman diagram  relevant for the dilepton search at the LHC and HL-LHC.}
        \label{feynmandiag}
    \end{minipage}
    \hfill
\end{figure}

Beyond the SM adventures typically invoke new Abelian gauge symmetries. These symmetries give rise to a new massive vector boson, a $Z^\prime$ field that can be searched for in current and future colliders \cite{Pisano:1992bxx,Alves:2023vig,Robinett:1982gw,ATLAS:2019erb, RamirezBarreto:2010vji, Cogollo:2020afo}. In our particular case, we are particularly interested in placing lower mass bounds on the B-L gauge symmetry \cite{Mohapatra:1980qe, Khalil:2006yi,Klasen:2016qux,Li:2010rb, Khalil:2008ps} using the LHC and HL-LHC (High-Luminosity LHC). In this case the gauge group reads $SU(3)_{C} \times SU(2)_{L} \times U(1)_{Y} \times U(1)_{B-L}$. This new gauge symmetry is anomalous and for this reason, three right-handed neutrinos are added to ensure gauge anomaly cancellation \cite{Arcadi:2023lwc}. The mass of this $Z^\prime$ field can be generated either via the spontaneous breaking of the B-L symmetry or by a Stueckelberg mechanism \cite{Ruegg:2003ps}. The presence of right-handed neutrinos allows us to easily implement the type I seesaw mechanism, which originally predicted right-handed neutrino masses at the GUT scale while generating active neutrino masses at the eV scale. We plan to probe the B-L symmetry via the $Z^\prime$ boson that can produced at the LHC via the production channel $pp \rightarrow {Z}^{\prime} \rightarrow \ell \ell$ as displayed in FIG.\ref{feynmandiag}. That said, the relevant Lagrangian for our analysis is,

\begin{equation*}
    \mathcal{L}\supset -\frac{1}{4}{F}^{\prime}_{\mu\nu}{F}^{'\mu\nu}+{g}_{B-L}{Q}_{X\chi}\overline{\chi}{\gamma}^{\mu}\chi{Z}^{\prime}_{\mu} 
\end{equation*}

\begin{align}
&&+{g}_{B-L}{Q}_{X\ell}\sum_{\ell=e,\mu,\tau} \overline{\ell} {\gamma}^{\mu} \ell {Z}^{\prime}_{\mu} + {g}_{B-L} {Q}_{Xq}\sum_{i=1,...,6} \overline{q_{i}} {\gamma}^{\mu} q_{i} {Z}^{\prime}_{\mu}+
\nonumber\\
&&+ \, g_{B-L} Q_{XN}\, \bar{N_i} \gamma_\mu \gamma_5 N_i {Z}^{\prime}_{\mu}
    \label{relevant_lagrangian}    
\end{align}where $Q_{X}$ are the respective B-L charge of the fermions as presented in TABLE \ref{table1}, $g_{B-L}$ is the coupling constant of the new $B-L$ symmetry group, and $F^{\prime}_{\mu\nu}$ is the new strength tensor.

\begin{table}[h]
\begin{tabular}{|c|c|c|c|c|c|}
\hline
Field & $SU(3)_{C}$ & $SU(2)_{L}$ & $U(1)_{Y}$ & $U(1)_{B-L}$ & ${Z}_{2}$ \\ \hline
 ${\ell}_{iL}$ & 1 & 2 & -1/2 & -1 & 1 \\ \hline
 ${\ell}_{iR}$ & 1 & 1 & -1 & -1 & 1 \\ \hline
 ${N}_{iR}$ & 1 & 1 & 0 & -1 & 1 \\ \hline
 $\chi$ & 1 & 1 & 0 & 1/3 & -1 \\ \hline
 ${q}_{iL}$ & 3 & 2 & 1/6 & 1/3 & 1 \\ \hline
 ${q}^{u}_{iR}$ & 3 & 1 & 2/3 & 1/3 & 1 \\ \hline
 ${q}^{d}_{iR}$ & 3 & 1 & -1/3 & 1/3 & 1 \\ \hline
 H & 1 & 2 & -1/2 & 0 & 1 \\ \hline
 ${\phi}_{s}$ & 1 & 1 & 0 & 2 & 1 \\ \hline

\end{tabular}
\caption{Particles and their respective charges under the $SU(3)_{C} \times SU(2)_{L} \times U(1)_{Y} \times U(1)_{B-L}$ group.}
\label{table1}
\end{table}

Using this framework, we are able to place tighter constraints on the B-L model. It's important to note that we will assume the right-handed neutrinos to be sufficiently heavier than the $Z^\prime$ field, $M_{N_R} \gg M_{Z^\prime}/2$. This means that right-handed neutrinos do not contribute to the $Z^\prime$  decay width—a scenario we will refer to as {\it base model}. In the {\it base model} the $Z^\prime$ decays exclusively into SM fermions. We emphasize that our assumption is motivated by the type I seesaw mechanism. We stress that assuming $M_{N_R}< M_Z^\prime/2$ would not yield meaningful changes because in the B-L model the $Z^\prime$ couples to all fermions proportionally to $g_{B-L}$. In order words, the addition of three light copies of right-handed neutrinos will bring meaningful changes to the branching ratio into charged leptons. 

However, to expand our theoretical framework, we also explore the effects of incorporating exotic fermions and/or dark matter particles, which may result in significant invisible decay channels. We investigate how these invisible decay modes might influence the mass bounds established for the $Z^\prime$ boson, thus providing a more comprehensive understanding of the potential implications and constraints within the B-L model.

In the next section, we describe the reasoning followed to obtain the limits using public data from the ATLAS collaboration.

\section{Strategy to extract constrains from the ATLAS data.}
\label{ATLASanalysis}

\begin{figure}[ht]
    \begin{minipage}[b]{0.5\textwidth}
        \includegraphics[width=\textwidth]{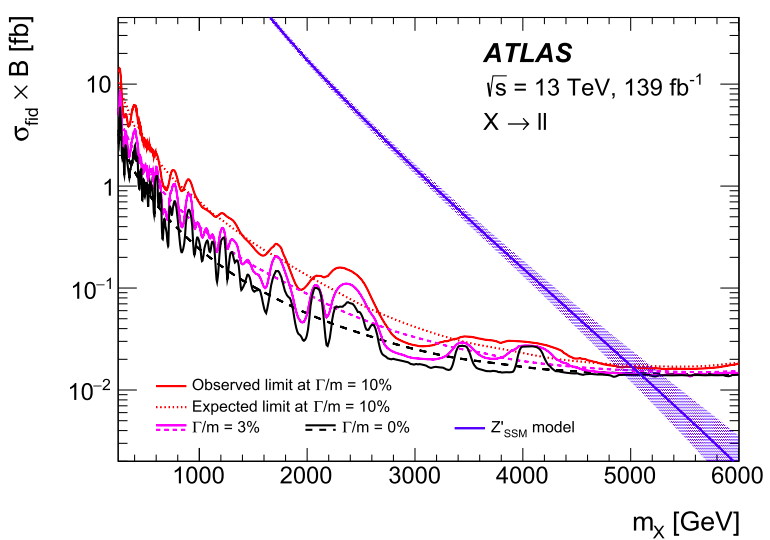}
        \caption{Experimental result vs theoretical prediction of the SSM model revealed by the ATLAS experiment \cite{ATLAS:2019erb}.}
        \label{atlas_result}
    \end{minipage}
    \hfill
\end{figure}

The ATLAS collaboration presents results related to the search for a new Spin-1 resonance using proton-proton ($pp$) collisions collected during LHC Run 2 at a center-of-mass energy of $\sqrt{s} = 13$ TeV and with an integrated luminosity of 139${fb}^{-1}$ \cite{ATLAS:2019erb}. The results, shown in Figure \ref{atlas_result}, comprise invariant mass measurements of dielectron and dimuon resonances in the energy range 250–6000 GeV.

\begin{figure}[ht]
    \begin{minipage}[b]{0.5\textwidth}
        \includegraphics[width=\textwidth]{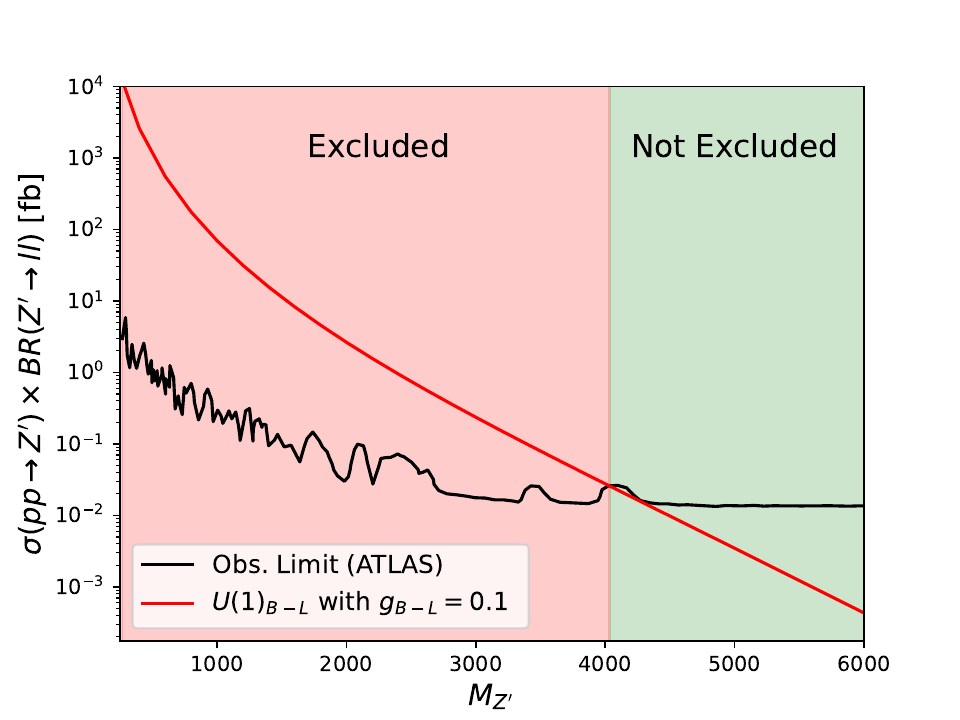}
        \caption{Excluded and non-excluded cross-sections of the $pp \to {Z}^{\prime} \to \ell\ell$ process in the B-L model with $g_{B-L}=0.1$.}
        \label{excluded_regions_01}
    \end{minipage}
    \hfill
\end{figure}

The collaboration defined event selection criteria for final states containing electrons and muons. Electron candidates were required to have a transverse energy $E_T > 30$ GeV and pseudorapidity $|\eta| < 2.47$. Muon candidates satisfied the same $E_T$ threshold but with $|\eta| < 2.5$. While these criteria ensured efficient data collection given the experimental setup, no excess above background was observed. The absence of a statistically significant deviation allowed the collaboration to set 95\% C.L. upper limits on the ${Z}^{\prime}$ mass for three models that contain a new spin-1 resonance. The main result published by ATLAS (FIG \ref{atlas_result}) show the experimental (black, pink and red lines) and theoretical prediction for $\sigma(pp \to Z') \times BR(Z' \to \ell\ell)$ as a function of the electron/positron invariant mass considering various models like Sequential Standard Model (violet line). To interpret these results and extract mass constraints on a hypothetical $Z^{\prime}$, we need to understand the relationship between the production cross section ($\sigma$) and the number of observed events ($N$): $N \propto \sigma L$. Specifically for this case, this can be translates to the hypothesis $N \propto \sigma(pp \to Z') \times BR(Z' \to \ell\ell) \times L$, where $L$ is the integrated luminosity of the experiment. Interpreting the cross section as a number of events observed, the experimental limits shown in FIG. \ref{atlas_result} represent the maximum number of lepton invariant mass events consistent with the selection criteria described above. Any BSM model predicting a cross section exceeding this limit is excluded since no excess was observed. This is illustrated for our case in FIG \ref{excluded_regions_01}, where, for example, setting $g_{B-L} = 0.1$ in the Lagrangian Eq. \eqref{relevant_lagrangian} divides the parameter space into excluded and allowed regions.

That said, we followed this procedure to obtain the ${Z}^{\prime}$ masses excluded for a given value of the coupling constant $g_{B-L}$. To do so, we implemented the model in the FeynRules package \cite{Alloul:2013bka} to carry out the Monte Carlo simulation using MadGraph \cite{Alwall:2014hca}. A scanning algorithm was codded to perform this task automatically, where a scan is performed over the parameter space ($g_{B-L}$,$M_{{Z}^{\prime}}$) considering two scenarios: (1) ${Z}^{\prime}$ boson decaying exclusively to SM fermions; (2) ${Z}^{\prime}$ boson decaying in SM fermions and invisible channels. The core of the algorithm employs a grid search to evaluate $\sigma(pp \to {Z}^{\prime}) \times BR({Z}^{\prime} \to \ell\ell)$ for various ($g_{B-L}$, $M_{Z'}$) combinations. The output is a uniformly distributed grid of parameter values where the calculated cross sections are then compared against the ATLAS upper limits for the two scenarios that will be presented below.  \newline

\section{Scenario 1: Without invisible decays}
\label{scenario1}

\begin{figure}[ht]
    \begin{minipage}[b]{0.5\textwidth}
        \includegraphics[width=\textwidth]{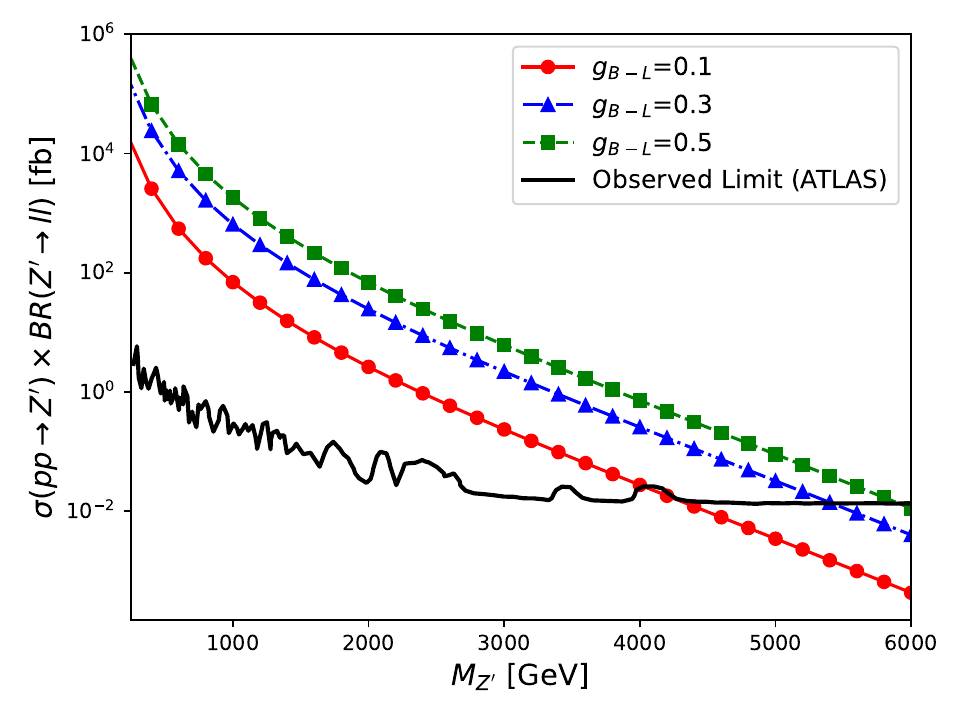}
        \caption{The production cross-section times the branching ratio into charged leptons ($ee,\mu\mu$) in the B-L model.}
        \label{three_cases_of_couplings}
    \end{minipage}
    \hfill
\end{figure}

In this scenario we analyze the {\it base model}, that is, no additional consideration was made on the Lagrangian of Eq. \ref{relevant_lagrangian} describing the interactions of the ${Z}^{\prime}$ boson. In this case, the ${Z}^{\prime}$ decays exclusively into SM particles \footnote{We assumed the right-handed neutrinos to be much heavier than the $Z^\prime boson$. This choice is justified because this B-L model features a type I seesaw mechanism, with right-handed neutrinos masses much larger than the TeV scale.}. FIG. \ref{three_cases_of_couplings} illustrates three examples of this first scenario (${g}_{B-L}=0.1, \ 0.3, \ 0.5$).

ATLAS data allows exploration of the coupling constant ${g}_{B-L}$ up to approximately 0.5 since no events were observed with dilepton invariant masses greater than 6 TeV. In order words, the experiment loses statistics for $M_{Z^\prime} > 6$~TeV.  In particular, for $M_{{Z}^{\prime}} = 6000$ GeV, less than one event is predicted. Therefore, we can only probe $g_{B-L}> 0.5$ if the ${Z}^{\prime}$ has new decay channels as we investigate in the next section. 

\begin{figure}[ht]
    \begin{minipage}[b]{0.5\textwidth}
        \includegraphics[width=\textwidth]{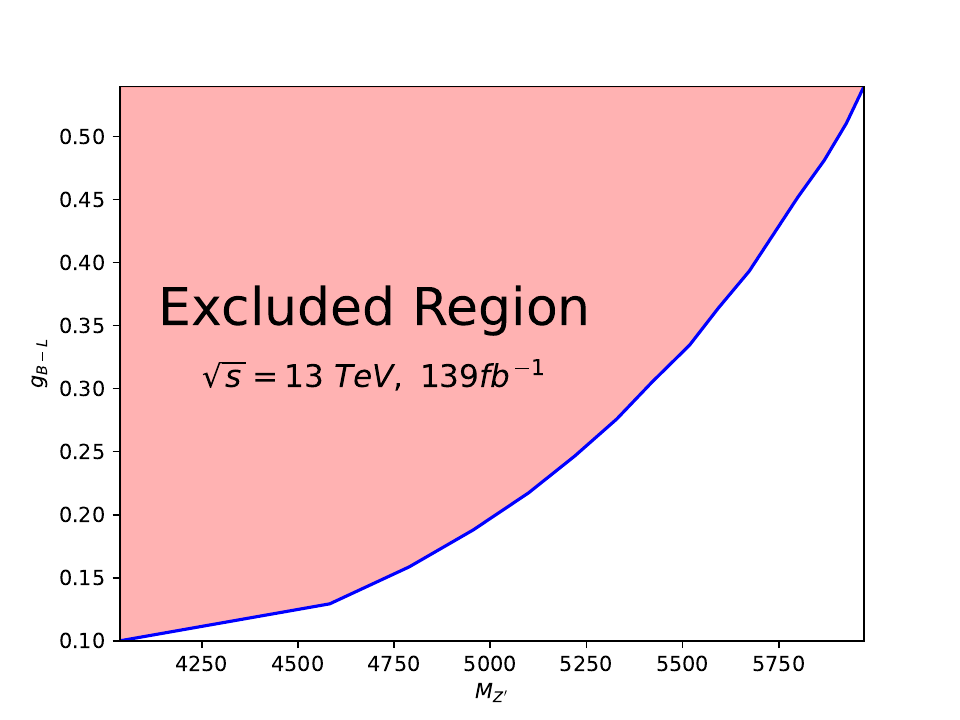}
        \caption{Exclusion limit based on ATLAS public data using with a center-of-mass energy of $\sqrt{13}$~TeV and an integrated luminosity of $\mathcal{L}=139 fb^{-1}$. Interestingly, we notice that LHC has surpassed the long-standing LEP bound regardless of the choice for the gauge coupling.}
        \label{excluded_param_BL}
    \end{minipage}
    \hfill
\end{figure}

For completeness, we use the scannning algorithm on Madgraph to cover a larger parameter space and derive (FIG. \ref{excluded_param_BL}) which shows the excluded parameter space of the original B-L model. We can now compare our findings with the good and old limit from LEP that reads, $M_{Z^\prime} > g_{B-L}\, 7\,{\rm TeV} $ \cite{Electroweak:2003ram,Elgammal:2024elw,ALEPH:2013dgf,ParticleDataGroup:2024cfk}. Taking $g_{B-L}=0.1$, LEP implies $M_{Z^\prime} > 700$~GeV, which is clearly weaker than LHC bound. If we consider $g_{B-L}=0.5$, LEP imposes $M_{Z^\prime} > 3.5$~TeV, whereas LHC enforcers  $M_{Z^\prime} > 6$~TeV. We have reached an era where LHC is more constraining than the long-standing LEP bounds regardless of the gauge coupling choice. \newline

\section{Scenario 2: With invisible decays}
\label{scenario2}

In the previous scenario, it was considered that the total width of the ${Z}^{\prime}$ boson encompassing fermions from the standard model and a small fraction of a new neutrino that is part of the model under consideration. In this second scenario, we rescale the production cross section ($\sigma (pp \rightarrow Z^\prime  \rightarrow \ell \ell$) to account for the invisible decay. Knowing the branching ratio into charged leptons ($ee,\mu\mu$) is given by,

\begin{equation}
    BR({Z}^{\prime} \to \ell\ell) \rightarrow (1-{\rm BR_{inv}})BR({Z}^{\prime} \to \ell\ell),
    \label{newcontributionequation1}
\end{equation}we can parameterize the production cross section in the narrow width approximation as, 
\begin{equation}
    \sigma(pp\rightarrow{Z}^{\prime}\rightarrow \ell\ell) \sim  \sigma(pp\rightarrow{Z}^{\prime})BR({Z}^{\prime} \to \ell\ell)
(1-{\rm BR_{inv}}),
    \label{newcontributionequation2}
\end{equation}where ${\rm BR_{inv}}$ represents the branching ratio into invisible decays. Note that if ${\rm BR_{inv}}=0$ we return the {\it base model}.

\begin{figure}[ht]
\centering
    \begin{minipage}[b]{0.5\textwidth}
        \includegraphics[width=\textwidth]{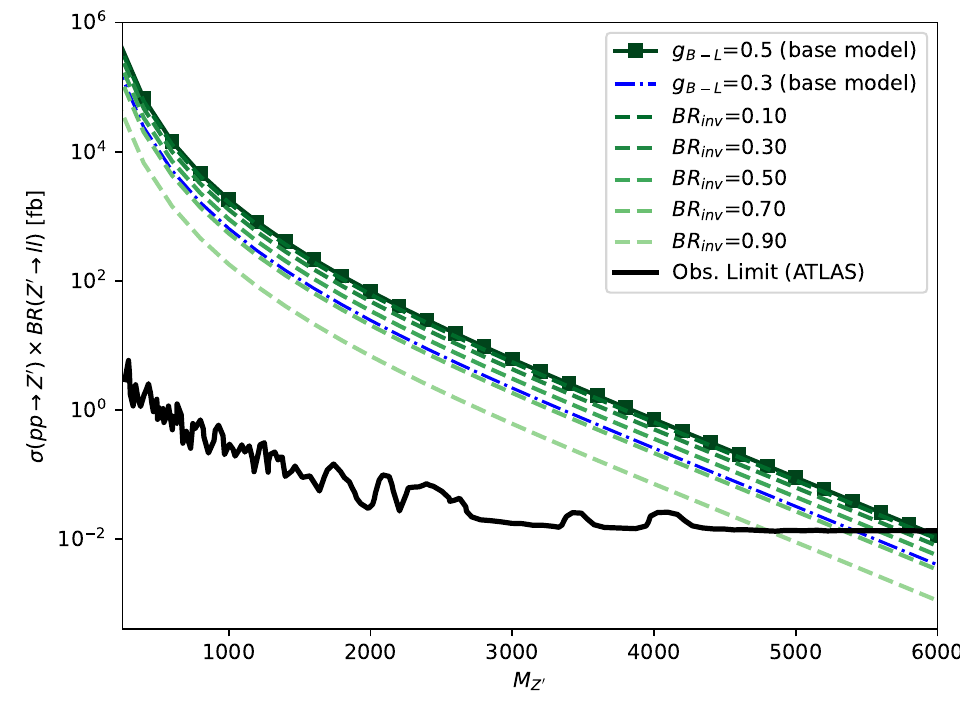}
        \caption{Production cross section $\sigma (pp \rightarrow Z^\prime) \times BR (Z^\prime \rightarrow \ell \ell)$ as a function of the $Z^\prime$ mass adopting $g_{B-L}=0.5$ and $g_{B-L}=0.3$, for several ${\rm BR_{inv}}=0.1-0.9$. }
        \label{crossinvdecays}
    \end{minipage}
\end{figure}

From Eqs. \ref{newcontributionequation1} and \ref{newcontributionequation2}, we easily conclude that as soon as we add invisible decays the $Z^\prime$ signal dwindles, and consequently the upper mass limit for the new ${Z}^{\prime}$ boson weakens as seen in FIG. \ref{crossinvdecays}. In fact, for $g_{B-L}=0.5$ the previous bound was $M_{Z^\prime} > 6$~TeV, but now reads $M_{Z^\prime} > 4.8$~TeV for ${\rm BR_{inv}}=0.9$. 

\begin{figure}[ht]
    \begin{minipage}[b]{0.5\textwidth}
        \includegraphics[width=\textwidth]{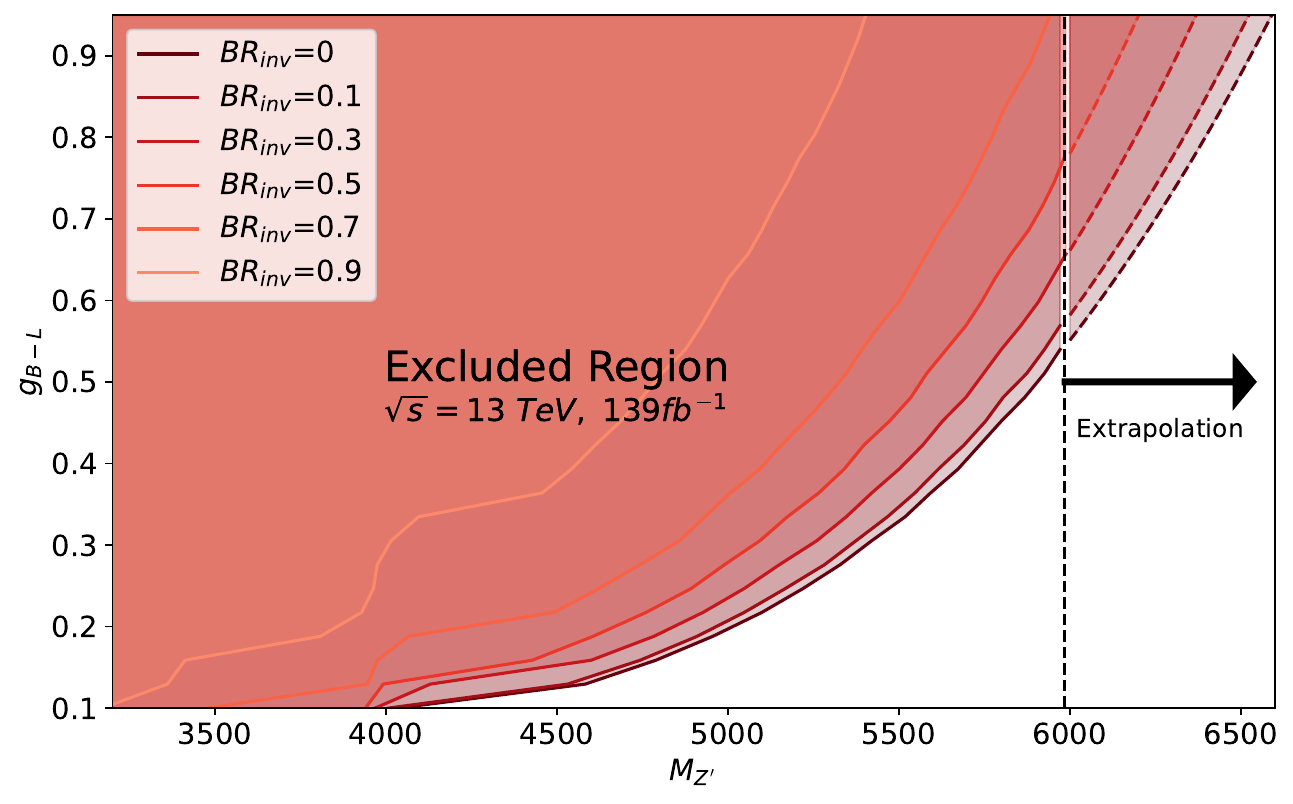}
        \caption{Similar to FIG.\ref{excluded_param_BL} but in the  presence of invisible decays. It is clear that invisible decays clearly weaken the limits. Nevertheless, the long-standing LEP bound is outperformed.}
        \label{excluded_param}
    \end{minipage}
    \hfill
\end{figure}

In a similar vein to the previous section, in FIG \ref{excluded_param} we exhibit the excluded parameter space ($g_{B-L} \times {M}_{{Z}^{\prime}}$) in the presence of invisible decays. 

It is important to point out that some benchmark models can become indistinguishable. In FIG. \ref{crossinvdecays} we notice that the $Z^\prime$ signal for ${g}_{B-L}=0.3$ and $BR_{inv}=0$ is equivalent to the setup with ${g}_{B-L}=0.5$ and $BR_{inv}=0.7$. Therefore, the search for heavy dilepton resonances are promising because the signal is clear, but once a signal is observed, it is tricky to unveil the underlying model. Therefore, the continuous search for $Z^\prime$ signals in different observables it is paramount. \newline

\section{HL-LHC Sensitivity}
\label{HLLHCsec}
To fully harness the capabilities of the LHC, an upgrade to increase its luminosity, along with substantial enhancements to the major experiments, has been sanctioned by the CERN Council. The High Luminosity LHC (HL-LHC) upgrade aims to accumulate an integrated luminosity of $3 ab^{-1}$ in proton-proton collisions at a center-of-mass energy of 14 TeV, thereby maximizing the potential of the LHC to discover new phenomena. That said, we estimate the HL-LHC reach using the collider reach code \cite{Thamm:2015zwa}. 

As we are dealing with searches for heavy dilepton resonances where the signal acceptance and efficiency are nearly independent of the resonance mass and center of mass energy, we can rely on the collider reach code to estimate projected sensitivities of future proton-proton colliders. Assuming the CTL8NNLO parton distribution function computed at the next-to-next leading order \cite{Hou:2019efy}, we find the forecasted bounds as summarized in Tables \ref{tab:HL-LHC1} and Table \ref{tab:HL-LHC2}.

\begin{table}[h]
{\bf HL-LHC projects on the B-L model for $Z^\prime$ visible decays}
\centering
\begin{tabular}{|c|c|}
\hline
  Coupling  & HL-LHC Projection  \\
\hline
  $g_{BL}=0.1$  & $M_{Z^\prime} > 5680$~GeV\\
  $g_{BL}=0.2$  & $M_{Z^\prime} > 6688$~GeV\\
   $g_{BL}=0.3$  & $M_{Z^\prime} > 7089 $~GeV\\
   $g_{BL}=0.4$  & $M_{Z^\prime} > 7364 $~GeV\\
    $g_{BL}=0.5$  & $M_{Z^\prime} > 7578$~GeV\\
   \hline
\end{tabular}
\caption {HL-LHC projected bounds for $\sqrt{s}=14$~TeV, $\mathcal{L}=3 ab^{-1}$ in the absence of new decay modes beyond the original B-L model.} \label{tab:HL-LHC1} 
\end{table}

\begin{table}[h]
{\bf HL-LHC projections on the B-L model for invisible decays}
\centering
\begin{tabular}{|c|c|c|}
\hline
 Coupling  & Branching & HL-LHC Projection  \\
 \hline
 \multirow{4}{*}{$g_{B-L}=0.2$} & $BR_{inv}=0.9$  & $M_{Z^\prime} > 5512$~GeV\\
              & $BR_{inv}=0.7$  & $M_{Z^\prime} > 5831$~GeV\\
              & $BR_{inv}=0.5$  & $M_{Z^\prime} > 6341$~GeV\\
              & $BR_{inv}=0.3$  & $M_{Z^\prime} > 6520$~GeV\\
 \cline{1-3}
 \multirow{4}{*}{$g_{B-L}=0.3$} & $BR_{inv}=0.9$  & $M_{Z^\prime} > 5667$~GeV\\
               & $BR_{inv}=0.7$  & $M_{Z^\prime} > 6500$~GeV\\
               & $BR_{inv}=0.5$ & $M_{Z^\prime} > 6745$~GeV\\
               & $BR_{inv}=0.3$ & $M_{Z^\prime} > 6922$~GeV\\
 \cline{1-3}
 \multirow{4}{*}{$g_{B-L}=0.4$} & $BR_{inv}=0.9$  & $M_{Z^\prime} > 6233$~GeV\\
               & $BR_{inv}=0.7$  & $M_{Z^\prime} > 6782$~GeV\\
               & $BR_{inv}=0.5$  & $M_{Z^\prime} > 7029$~GeV\\  
               & $BR_{inv}=0.3$  & $M_{Z^\prime} > 7195$~GeV\\ 
 \cline{1-3}
 \multirow{4}{*}{$g_{B-L}=0.5$} & $BR_{inv}=0.9$  & $M_{Z^\prime} > 6473$~GeV\\    
               & $BR_{inv}=0.7$  & $M_{Z^\prime} > 7003$~GeV\\    
               & $BR_{inv}=0.5$  & $M_{Z^\prime} > 7245$~GeV\\ 
               & $BR_{inv}=0.3$  & $M_{Z^\prime} > 7403$~GeV\\
 \cline{1-3}
 \multirow{4}{*}{$g_{B-L}=0.6$} & $BR_{inv}=0.9$  & $M_{Z^\prime} > 6642$~GeV\\
               & $BR_{inv}=0.7$  & $M_{Z^\prime} > 7182$~GeV\\
               & $BR_{inv}=0.5$  & $M_{Z^\prime} > 7416$~GeV\\ 
               & $BR_{inv}=0.3$  & $M_{Z^\prime} > 7573$~GeV\\ 
 \hline
\end{tabular}
\caption{HL-LHC constraints for $\sqrt{s}=14{\rm TeV}$, $\mathcal{L}=3 ab^{-1}$ in the presence of new invisible decays.} 
\label{tab:HL-LHC2} 
\end{table}

From the Tables \ref{tab:HL-LHC1}-\ref{tab:HL-LHC2} we see that HL-LHC can reach masses above $6$~TeV even $BR_{inv}=0.9$. Therefore, the HL-LHC will indeed set new standards for new resonance searches. 

\section{Conclusions}
\label{conclusions}

Some of the open problems of the SM such as dark matter and neutrino masses are often addressed in the beyond the SM theories featuring a $U(1)_{B-L}$ gauge symmetry.  These models feature a massive $Z^\prime$ that has sizeable couplings to quarks and leptons. 
Leveraging publicly available dilepton data from the ATLAS collaboration, collected during LHC run 2 at a center-of-mass energy of $\sqrt{s}=13$~TeV and an integrated luminosity of $\mathcal{L}=139 fb^{-1}$, we have updated the lower mass bounds for the $Z^\prime$ boson associated with this symmetry. Our analysis also extends to variants of the B-L model that include significant invisible decay channels, potentially due to dark matter particles or exotic neutral leptons.

We have established lower mass bounds across a broad swath of the B-L parameter space. These quantitative findings are detailed in FIG. \ref{excluded_param_BL} and FIG. \ref{excluded_param}. Furthermore, anticipating the capabilities of the upcoming High Luminosity LHC (HL-LHC), which will operate at $\sqrt{s}=14$~TeV and $\mathcal{L}=3 ab^{-1}$, we forecasted HL-LHC constraints as summarized in TABLE \ref{tab:HL-LHC1}-\ref{tab:HL-LHC2}.

In conclusion, our work marks a new milestone in the search for B-L symmetry, with the LHC surpassing the long-standing bound set by LEP even with the inclusion of sizeable invisible decays. 

\section*{Acknowledgements}

We thank Yara Coutinho, Gabriela Hoff and Bertrand Laforge for discussions. This work was funded by ANID - Millennium Science Initiative Program - ICN2019\_044. J. Zamora-Saa was partially supported by FONDECYT grant 1240216 and 1240066. FSQ is supported by Simons Foundation (Award Number:1023171-RC), FAPESP Grant 2018/25225-9, 2021/01089-1, 2023/01197-4, ICTP-SAIFR FAPESP Grants 2021/14335-0, CNPq Grant 307130/2021-5, PROPESq-UFRN grant 758/2023, and IIF-FINEP grant 213/2024.

\bibliography{referencias}

\end{document}